\documentclass[%
 reprint,
 amsmath,amssymb,
 aps,
]{revtex4-1}
\usepackage{amsmath}
\usepackage{graphicx}
\usepackage{enumitem}
\begin{document}

\newcommand{\refeq}[1]{Eq.~(\ref{#1})}
\newcommand{\reffig}[1]{Fig.~\ref{#1}}

\title{New Wrinkles on an Old Model: \\{\small Correlation Between Liquid Drop Parameters and Curvature Term}}

\author{L. G. Moretto}
\author{P. T. Lake}
\author{L. Phair}
\affiliation{
Lawrence Berkeley Laboratory, One Cyclotron Road, Berkeley, CA 94720
}

\author{J. B. Elliott}
\affiliation{
Lawrence Livermore Laboratory, 7000 East Avenue, Livermore, CA 94550
}

\date{\today}

\begin{abstract}
The relationship between the volume and surface energy coefficients in the liquid drop $A^{-1/3}$ expansion of nuclear masses is discussed. The volume and surface coefficients share the same physical origin and their physical connection is used to extend the expansion with a curvature term.  A possible generalization of the Wigner term is also suggested.  This connection between coefficients is used to fit the experimental nuclear masses.  The excellent fit obtained with a smaller number of parameters validates the assumed physical connections and the usefulness of the curvature term.
\end{abstract}

\maketitle

{\bf Introduction.}  Nuclear masses and their dependence on atomic and mass number gave essential information about the nature of nuclear forces.  They also led to the formulation of the liquid drop model, arguably the most precise and easily interpretable description of the masses themselves \cite{bib1}.

Empirical trends and scientific intuition led to the formulation of the liquid drop model.  In its traditional form, the liquid drop model approximates the binding energy of a given nucleus of mass number $A$ and charge $Z$ as \cite{bib1}:
\begin{align}
E_B(A,Z)=-a_v A +a_s A^{2/3} + a_c \frac{Z(Z-1)}{A^{1/3}} \nonumber\\
+ a_a \frac{(A-2Z)^2}{A} \pm \frac{\delta}{\sqrt{A}}.
\label{eq1a}
\end{align}
The five terms in this equation are associated with five independent aspects of nuclei expected to affect the binding energy.  These aspects are the nuclear volume, surface, Coulomb repulsion, proton-neutron asymmetry, and pairing.  A fit of this equation to nuclear masses gives the coefficients and reproduces the experimental values to within 1\% or $\sim$10 MeV for heavy nuclei.

This is an outstanding result that attests to the profound physical content of the overall equation and to the interpretation of its individual terms.  The residual 1\% discrepancy is due to shell structure.  The shell corrections, evaluated according to the Strutinsky procedure \cite{bib2} and grafted onto the liquid drop model, permit an accurate evaluation of nuclear masses and fission barriers to within 1-2 MeV \cite{bib3,bib15,bib4,bib5}.  This hybrid approach remains to this day the yet unmatched paragon for more sophisticated models such as Hartree-Fock-Bogoliubov \cite{bib11,bib12}.

Motivated by this early success of the liquid drop model, many additional terms have been suggested, each with its own physical interpretation.  An example of this is found in Myers and Swiatecki \cite{bib3} who suggested:
\begin{align}
E_B(A,Z)= -a_v &\left(1-k\frac{I^2}{A^2}\right) A +a_s \left(1-k\frac{I^2}{A^2}\right) A^{2/3} \nonumber\\
&+ a_c \frac{Z(Z-1)}{A^{1/3}} + W\frac{|I|}{A} - C_4\frac{Z^2}{A}\pm \frac{\delta}{\sqrt{A}},
\label{eq1}
\end{align}
where $I{=}A-2Z$.  The main difference between \refeq{eq1a} and \refeq{eq1} is the extension of the neutron-proton asymmetry to the surface energy term.  Also, a term linear in $|I|$ was introduced.

The further physical insight found in writing the asymmetry energy as in \refeq{eq1} is the connection it implies between volume and surface energies.  The authors argued that the change of the volume energy due to the neutron-proton asymmetry $I$ should be reflected in the surface energy of the system as well, though stating that this was done without empirical evidence \cite{bib3}.  The natural implication is that the surface and volume energies are related through their common origin.

A term linear in $|I|$ was originally suggested by Wigner in considering the exchange force of nucleons \cite{bib9}.  An empirical observation of such a dependency in the masses was reported by Myers and Swiatecki, hence its addition in the above equation \cite{bib3}.

The last alteration was the addition of a term proportional to $Z^2/A$, which is to account for the difference in the Coulomb energy due to the diffuse nuclear surface.  Again, this trend was not observed in the nuclear data, but was added because it would be anticipated.  We will not discuss the role of this term in the nuclear binding since it requires an additional fitting parameter to describe what is ambiguous in the data.

In this paper we extend and generalize the insights discussed above.  In particular, we argue that the relation between volume energy and the surface energy is strong enough that their coefficients should not be taken as independent variables.  Furthermore, we discuss the need of a third term arising from the same physics, proportional to $A^{1/3}$, in order to create a consistent physical picture of the nuclear binding energy.  Finally, a linear term in $|I|$ is naturally introduced when treating the asymmetry term as the expectation value of the isospin, $T^2$.

A revised liquid drop model is fit to the experimental binding energies of the nuclides to test these considerations.  The importance of the revised terms is assessed by comparing the fit of the original and revised models.

{\bf The Liquid Drop Formula: A Truncated Series Expansion.}  The first term of \refeq{eq1} is aptly called the volume term.  Its proportionality to $A$ indicates saturating forces leading to constant density and binding energy per nucleon.  The obvious similarity to molecular fluids led naturally to the introduction of the second term, the surface term.  Its proportionality to $A^{2/3}$ speaks to the lack of saturation on the nuclear surface, whose area, through the constant density of the fluid, should indeed be proportional to $A^{2/3}$.  Progressing along the same line, it was widely appreciated that the surface term is a finite size correction and that additional terms in the expansion might be needed, such as a curvature term.

Generally, we can think of a generalized liquid drop formula as a rapidly converging series expansion in powers of $A^{-1/3}$, known as the leptodermous expansion \cite{bib14}:
\begin{equation}
E_B = -a_v A + a_s A^{2/3} + a_r A^{1/3} + ...
\label{eq2}
\end{equation}
It is left to be determined how many terms in the expansion are necessary to describe the physics of the nuclear system.  The incorporation of a curvature term, with its coefficient $a_r$, proportional to $A^{1/3}$ is almost demanded by the truly small size of nuclei ($A \leq 300$) compared to the size of the drops typically considered in molecular fluids, such as aerosols, where $A \geq 10^6$.  Higher order terms also may be of importance due to the small size of nuclei, but would be intractable without an understanding of the lower order curvature term.

The role of the curvature term in nuclear systems was considered only recently and has yielded ambiguous results \cite{bib3,bib15,bib4,bib5}.  The increased number of parameters and the ability of the traditional liquid drop formula without curvature to fit the data made the problem of identifying the magnitude of this term rather difficult.  We believe that it is possible to shed additional light on this subject by considering the physical origin of the various terms.

Volume and surface terms both arise from the same physical property of nuclear forces: saturation, and the lack thereof.  Thus, surface and volume terms should be related to one another, being themselves different effects of the same cause.  Furthermore, the experimental surface and volume coefficients turn out to be approximately equal.  Is this an accident or could they possibly be equal?

To answer this question, consider a system of small sticky cubes used to build larger, composite cubes.  These cubes interact only when in direct contact.  The system is characterized by some bond strength, $\epsilon$, when two faces are touching.  The energy of a cube of $A$ constituents is equal to a volume energy minus a surface energy, just as in the nuclear case.  Counting the number of bonds in a cube of size $A$ reveals:
\begin{equation}
E_B^{(cube)}(A)=-3A\epsilon + 3A^{2/3}\epsilon.
\end{equation}
Thus, in this model the volume and surface energy coefficients are {\it exactly} equal with $a_v{=}a_s{=}3\epsilon$.  Even though this is a simplified model in comparison to a nucleus, it exemplifies the fact that the volume and surface terms are strongly connected.  This insight motivates setting $a_v{=}a_s$ without any loss of information.

\begin{table}
\caption[]{Fits of the nuclear masses with \refeq{eq5} using different mass ranges and setting $a_r{=}0$.  All the parameters in units of MeV.  The value in the parentheses is the uncertainty in the last digit.}
\label{tab1}
\begin{ruledtabular}
\begin{tabular}{ c c c c c c }
Masses  &    $a_v$ &      $a_s$ &  $k$      & $a_c$       & $\delta$ \\
\colrule
 50-100 & 15.39(4) & 16.81(10) & 1.742(7)  & 0.686(3)    & 10.3(5)  \\
100-150 & 15.39(2) & 16.68(7)  & 1.771(3)  & 0.6917(14)  & 12.4(3)  \\
150-200 & 15.11(2) & 15.66(8)  & 1.748(3)  & 0.6760(12)  & 13.5(3)  \\
200-250 & 15.18(6) & 15.7(2)   & 1.768(5)  & 0.686(3)    & 13.3(4)  \\
\end{tabular}
\end{ruledtabular}
\end{table}

One difference between this simple model system and a nucleus is the diffuseness of the nuclear surface.  What effect does a diffuse surface have on the binding energy of a drop?  Since the volume energy is a property of the bulk system it would remain unchanged.  The fact that the system naturally becomes diffuse means that it gains a larger binding energy in doing so.  The surface energy would then be {\it lowered} in comparison to the sharp surface system.  This implies that the surface energy coefficient should be equal to or smaller than the volume energy coefficient, contrary to what is observed in traditional liquid drop fits to the nuclear masses.

As the system is made smaller, more terms in the leptodermous expansion may be needed to properly predict binding energies.  If one were to fit the expansion with an insufficient number of terms, what ailments would be observed?  The terms included in the equation would have to change from their nominal values to accommodate the lack of higher order terms.  Also, the deviation from the nominal value would be worse for smaller masses, where the higher order terms are more important.

As an example, consider nuclear binding energies in various mass ranges.  Each mass range can be fit  with \refeq{eq5}, using the fitting procedure that is described in the following section.  Table \ref{tab1} shows the results of such an exercise.  Most terms do not vary systematically as the mass range is changed, their variation being of the order of 1\%.  The exceptions are the surface energy and the pairing energy.  The pairing energy is of unrelated physics and is not discussed here.  The surface energy coefficient {\it decreases} as the mass range is incremented.  This trend indicates that the $A^{2/3}$ term is not sufficient in describing the lack of saturation in the system.  As the masses used in the fit increase, the surface term tends to the value of the volume coefficient.  Hence, both the need of a curvature term and setting $a_v{=}a_s$ are motivated.

\begin{figure}
\vspace{-1.5cm}
\begin{center}
\hspace*{-0.3cm}
\includegraphics[width=9.5cm] {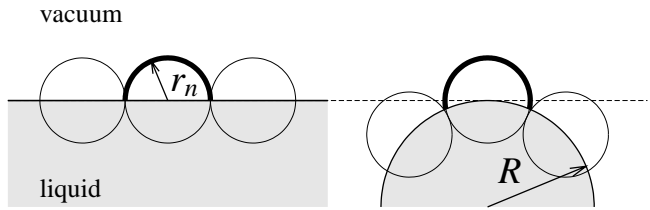}
\end{center}
\vspace{-2.75cm}
\caption[]{Schematic representation of the surface energy.  The system on the left represents a flat surface of an infinite liquid and the system on the right is a finite liquid drop.  The surface area of a constituent of radius $r_n$ exposed on the surface of a liquid drop of radius $R$ is more than that of a surface particle at the flat surface, as emphasized by the bold curve.}
\label{fig1}
\end{figure}

Now that we see the need for a curvature term, one may ask:  What is its origin?  To answer, let us consider a simple liquid with spherical molecules of radius $r_n$.  As shown on the left side of \reffig{fig1}, on a flat liquid surface the molecules should protrude half way on average, losing half of their binding energy.  If the liquid surface is curved like that of a sphere, as in the right side of \reffig{fig1} with a drop radius of $R$, the molecules protrude more, losing additional binding.  Thus, the curvature and surface terms arise from the same physical effect and their coefficients should be related.

In order to obtain a quantitative estimate of this effect, consider a model that is geometric in nature.  The surface energy is considered to be proportional to the protruding surface area of a constituent residing on the surface times the number present on the surface.  As a function of nuclear radius $R$, the resulting exposed surface area $S$ of a constituent on the surface is:
\begin{equation}
S=2\pi r_n^2\left(1+\frac{r_n}{2R}\right),
\end{equation}
with the limiting case of a planar system, $S{=}2\pi r_n^2$.  The number of particles on the nuclear surface is proportional to $A^{2/3}$.  The overall surface energy is then:
\begin{equation}
E_s = a_s A^{2/3} \left(1+\frac{r_n}{2R}\right).
\label{eq3a}
\end{equation}
Since nuclei exhibit a saturation density, the nuclear radius is approximated as $R{=}r_0A^{1/3}$, with $r_0$ being a constant.  Inserting this relation into \refeq{eq3a} yields:
\begin{equation}
E_s = a_s A^{2/3} + a_s \frac{r_n}{2r_0}A^{1/3}.
\label{eq3}
\end{equation}
Here we identify the usual surface term proportional to $A^{2/3}$.  Furthermore, we notice a curvature term proportional to $A^{1/3}$ with a coefficient that is dependent on the surface energy coefficient and the ratio of the ``molecule'' radius to $r_0$, which is directly related to the saturation density.

The above equation is reminiscent of the Tolman correction to the surface tension \cite{bib10}.  This term can be interpreted as the Tolman correction for the nuclear system in its ground state.

Naturally, deviations from sphericity of the molecules would involve a (temperature dependent) reorientation on the surface.  This would alter the simple relationship between volume, surface and curvature energies.  We will limit the discussion to the case of an isotropic force for the model presented here.

We may check the model further by putting experimental values into \refeq{eq3}.  Taking $r_0 \simeq 1.2$ fm \cite{bib4} and the radius of a free nucleon to be $r_n \simeq 0.9$ fm \cite{bib6}, yields:
\begin{equation}
a_r \simeq a_s \frac{0.9}{2.4} \simeq \frac{3}{8}a_s.
\label{eq4}
\end{equation}
These geometric arguments thus give a first order approximation as to the sign and magnitude of the curvature term.  Other aspects might influence the actual value of the curvature term in the nuclear system, but it would be notable if a proper fit to nuclear masses were to produce a value close to the above estimate.

To further appreciate the significance of the relation between volume, surface and curvature energies, consider the following.  What information is gained in knowing the leptodermous expansion for an arbitrary liquid in its ground state?

First, consider the volume energy.  The volume energy gives no information of the internal structure of the system.  It is just the scale which sets the size of the rest of the terms in the leptodermous expansion.

Now a measurement of the system's surface energy is made.  The {\it particle density} of the system can be deduced by comparing the surface and volume energies.  This is done by anticipating that the two coefficients will be the same in terms of $A$ and $A^{2/3}$, respectively.  Avogadro's number could thus be inferred.

Finally, the curvature energy is determined and from it the {\it size} of a single particle in the liquid can be estimated.  This is shown in \refeq{eq3}.

Here we see how the hierarchy of terms in the leptodermous expansion can be related to the internal structure of a fluid.  Even though this exercise is pedagogical in nature, it demonstrates the physical significance of each term.  It could have allowed Democritus to prove his atomic theory, had he been inclined to do so.

\begin{table*}
\caption[]{Fits from the four different mass equations as described in the text.  All parameters are in units of MeV.  The value in the parentheses is the uncertainty in the last digit.}
\label{tab2}
\begin{ruledtabular}
\begin{tabular}{ c c c c c c c c c }
Fit &    $a_v$ &      $a_s$ &  $a_r$   & $k$        & $a_c$       & $\delta$ & $r_n$ (fm)   & $\chi^2$ \\
\hline
A   & 15.597(7) & 17.32(2)  & $a_r{=}0$  &  1.8048(9) &  0.7060(4)  & 11.4(2) & ---       &  0.58    \\
B   & 14.843(3) & $a_v{=}a_s$  & $a_r{=}0$  & 1.7196(16) &  0.6585(4)  & 10.1(6) & ---      &  4.24    \\
C   & 15.25(3)  & 15.17(17) & 3.8(3)   &  1.779(2)  &  0.6932(11) & 11.3(2) & 0.60(5)   &  0.54    \\
D   & 15.264(4) & $a_v{=}a_s$  & 3.60(3)  & 1.7805(8) &  0.6938(3)   & 11.3(2) & 0.566(5) &  0.54    \\
\end{tabular}
\end{ruledtabular}
\end{table*}

{\bf Nuclear Mass Fit Results.}  We use a set of 2076 masses, corrected for microscopic effects according to M\"oller {\it et al.} \cite{bib7}.  These microscopic corrections account for the shell effects along with the effects associated with nuclear deformation.  The masses considered in the fits correspond to nuclear masses from reference \cite{bib8} with $N>7$, $Z>7$, and with experimental uncertainties less than 150 keV.  The lower limit of neutron and proton numbers is chosen to ensure that the included nuclei are large enough to be considered as liquid drops.  The restriction on the experimental uncertainties is not only due to the error of the mass, but also to the reliability of the shell correction for masses far away from stability.  The binding energy, $E_B$, of each nucleus is defined as:
\begin{equation}
E_B(A,Z)=Z m_p + (A-Z) m_n - M(A,Z) + \Delta_{shell}(A,Z),
\end{equation}
with $m_p$ and $m_n$ being the mass of a proton and neutron, respectively, $M$ is the experimental mass of the nucleus, and $\Delta_{shell}$ is the shell correction.  The liquid drop formula is fit to this binding energy with each nucleus given an equal weight.  The mean square deviation of the fit is used to evaluate its goodness:
\begin{equation}
\chi^2=\frac{\sum(E_i^{(ex)}-E_i^{(th)})^2}{N}.
\end{equation}

We use the following liquid drop formula:
%\begin{widetext}
%\begin{equation}
%%E_B = -a_v A \left(1-k\left(\frac{|I|(|I|+2)}{A^2}\right)\right) + a_s A^{2/3} \left(1-k\left(\frac{|I|(|I|+2)}{A^2}\right)\right) + a_r A^{1/3} \left(1-k\left(\frac{|I|(|I|+2)}{A^2}\right)\right) + a_c \frac{Z(Z-1)}{A^{1/3}} \pm \frac{\delta}{\sqrt{A}},
%E_B = (-a_v A + a_s A^{2/3} + a_r A^{1/3})\left(1-k\left(\frac{|I|(|I|+2)}{A^2}\right)\right) + a_c \frac{Z(Z-1)}{A^{1/3}} \pm \frac{\delta}{\sqrt{A}},
%\label{eq5}
%\end{equation}
%\end{widetext}
\begin{align}
E_B =& (-a_v A + a_s A^{2/3} + a_r A^{1/3})\left(1-k\left(\frac{|I|(|I|+2)}{A^2}\right)\right) \nonumber\\
&+ a_c \frac{Z(Z-1)}{A^{1/3}} \pm \frac{\delta}{\sqrt{A}},
\label{eq5}
\end{align}
where we insert the mass asymmetry dependence $I{=}A-2Z$ both in the volume and surface terms according to Myers and Swiatecki \cite{bib3}.  If the mass asymmetry term is interpreted as an ``isospin'' dependence, the term linear with $I^2$ should be treated as $T^2$, with $T{=}|I|/2$.  This ``isospin'' presents itself as the square $T^2$, which we rewrite (with a possibly unjustified quantal sensitivity) as $\langle T^2 \rangle {=} T(T+1) {=} |I|(|I|+2)/4$.  This introduces a linear term in $|I|$ without the addition of a new parameter, as opposed to a freely varying Wigner term \cite{bib9}.

\begin{figure}
\begin{center}
\includegraphics[width=8.5cm]{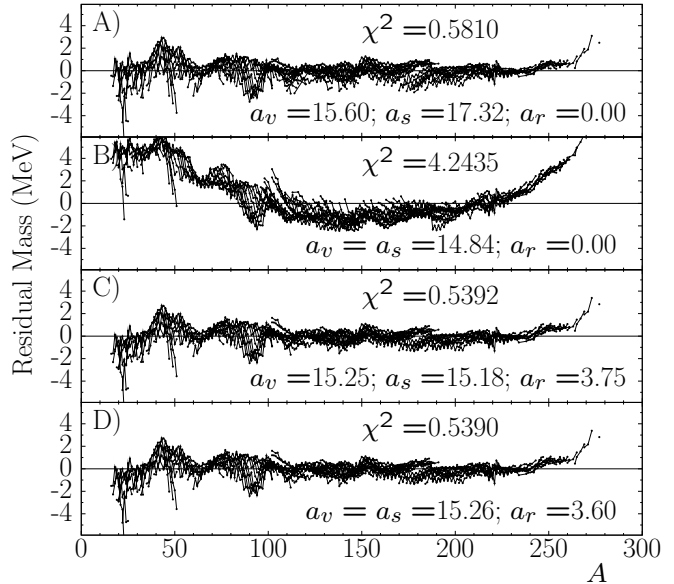}
\end{center}
\vspace{-0.5cm}
\caption[]{The residual mass from the corresponding fits.  The label in the top left corner of each plot corresponds to the fits listed in the text.  The connected lines represent chains of isotopes.}
\label{fig2}
\end{figure}

The following fits are performed:
\begin{enumerate}[itemsep=0pt,parsep=0pt]
\item[A.] $a_v$ and $a_s$ vary independently without a curvature term.
\item[B.] Same as above, but forcing $a_v{=}a_s$.
\item[C.] $a_v$ and $a_s$ vary independently with a curvature term.
\item[D.] Same as above, but forcing $a_v{=}a_s$.
\end{enumerate}
The Coulomb, mass asymmetry and pairing coefficients are left as free parameters in all of the above fits.  The results are shown in Table \ref{tab2} and are discussed below.  \reffig{fig2} shows plots of the residual masses of the fits, the exact binding energy with shell corrections included minus the binding energy predicted from the fitted formula.

Comparing fits A and B shows that setting $a_v{=}a_s$ without the curvature term does not ameliorate the situation. Quite to the contrary, the $\chi^2$ value is 8 times larger and the plot of the residual masses shows clear deviations.  Left without constraint, the surface term incorporates the curvature effects and becomes larger.

Comparing fits A and C, we observe that the introduction of the curvature term as a free parameter improves the resulting fits as expected.  But is the value of $a_r$ physically meaningful and how does it compare to the expectations of the geometric model?

Rearranging \refeq{eq4} gives the radius of the nucleon as
\begin{equation}
r_n=2.4\frac{a_r}{a_s},
\end{equation}
in units of fm.  Using this equation, the nucleon radius is found to be 0.60(5) fm, smaller than the experimental value of 0.84 fm \cite{bib6}.  This size of deviation is not unexpected from the crude approximations used, and it remains impressive that both the sign and relative magnitude are predicted.  Furthermore, the surface energy coefficient moves within error of the volume energy coefficient.  The other parameters change within 2\% between the two fits, showing consistent results.

By forcing $a_v{=}a_s$ with the presence of the curvature correction, as in fit D, the $\chi^2$ changes by a fraction of a percent.  Also, the parameters not associated with the saturating nuclear force are left unchanged.  Thus, no physics is lost with setting $a_v{=}a_s$.

Without taking into account the curvature term, the volume and surface parameters will tend to be irreconcilably different to be considered equal.  This explains the reason why the two terms have previously been treated as independent values.  The addition of the curvature term corrects this discrepancy, and it is found that the surface and volume energies are close to being equal, giving no visible difference in the fitting of the experimental data.

Another fit was performed using $\langle I^2 \rangle {=} |I|(|I|+x)$, with the added fit parameter $x$.  This addition is equivalent to introducing an adjustable Wigner term linear in isospin.  Table \ref{tab3} shows the fit with and without letting $x$ vary.  None of the other fit parameters change substantially.  As for $x$ itself, it is found to be 1.51(3), which slightly lowers the $\chi^2$ of the fit.  When written in the form presented by Myers and Swiatecki\cite{bib3}, this corresponds to a congruence energy of 41.3(8) MeV, which agrees with the value 42 MeV which they report.  With most of the parameters changing less than 1\%, the same physics is still captured by setting $\langle I^2 \rangle {=} |I|(|I|+2)$.

\begin{table}
\caption[]{Fits of the nuclear masses to the liquid drop model using different isospin dependencies.  The first sets $\langle I^2 \rangle {=} |I|(|I|+2)$, where as the second represents a fit to $\langle I^2 \rangle {=} |I|(|I|+x)$. All parameters are in units of MeV.  The value in the parentheses is the uncertainty in the last digit.}
\label{tab3}
\begin{ruledtabular}
\begin{tabular}{ c c c c c c c c }
    $a_v$ &      $a_r$ &  $k$        & $x$     & $a_c$      & $\delta$ & $\chi^2$ \\
\hline
15.264(4) &   3.60(3)  & 1.7805(8)  &   2     &  0.6938(3) & 11.3(2)  & 0.54  \\
15.247(4) &   3.76(3)  & 1.7944(10) & 1.51(3) &  0.6913(3) & 11.3(2)  & 0.46  \\
\end{tabular}
\end{ruledtabular}
\end{table}

%\begin{figure*}
%\begin{center}
%\includegraphics[width=\textwidth]{mass_all}
%\end{center}
%\vspace{-0.5cm}
%\caption[]{The residual mass from the corresponding fits.  The label in the top left corner of each plot corresponds to the fits listed in the text.  The connected lines represent chains of isotopes.}
%\label{fig2}
%\end{figure*}

{\bf Implications of the curvature term.}  The existence of a curvature energy, especially important in light nuclei, may imply effects hitherto undiscovered.  We give here two examples.

The curvature of the surface in the nuclear deformation landscape, and in particular at the fission saddle point, exhibits large variations going from positive to negative.  Therefore, the prediction of fission saddle point configurations and masses will be affected by the presence of a curvature term, which will acquire a tensorial form.

The fragment distribution predicted by the Fisher model \cite{bib13} is dependent on the surface energy of the clusters.  The theory uses a term proportional to $A^\sigma$ for this purpose.  Since the fragment yields are weighted heavily towards lighter fragments away from the critical temperature, the introduction of a curvature term would seem imperative.  Thus, the curvature term could alter predictions of the critical temperature in an unknown way.

{\bf Conclusion.}  Previous efforts have addressed the need of a curvature term in the liquid drop expansion of nuclear masses, but no consistent interpretation was made.  Some works state that it is unnecessary, and that it is enough to stop the expansion at the level of a surface term \cite{bib3}.  Other studies give conflicting results, and even the sign of the curvature correction remains ambiguous \cite{bib15,bib4,bib5}.  Some of these references do give results that agree with the ones here, but do not offer a physical picture. 

We demonstrate that the surface energy coefficient in the traditional liquid drop formula changes when different mass ranges are considered.  The decreasing trend in the surface energy coefficient with increasing mass number is consistent with the presence of a curvature term.  We present a consistent description of the curvature term's nature, determine its sign and demonstrate its presence in the nuclear masses.

Simple physical arguments predict that the volume and surface energy coefficients should be equal.  Without the introduction of the curvature term, the volume and surface energy coefficients appear to differ from each other.  With the addition of the curvature term, the two coefficients agree within error.

The nature of the ``Wigner'' term linear with isospin is also considered.  A slight change in the definition of the squared isospin, possibly quantum mechanical in nature, captures its relative magnitude without introducing an additional parameter.

What is gained through these considerations is a streamlined physical picture of the liquid drop model.  Consider the difference of the original liquid drop model in \refeq{eq1a} to the final equation presented here in \refeq{eq5}.  Even though the latter appears more complicated, there are the same number of free fit parameters as the former.  Instead of adding more and more terms to produce more and more exact representations of the nuclear masses, we have added a geometric physical picture and kept the same number of variables to obtain a more accurate result.  The lessons learned with this equation are more telling than letting all the parameters free.

{\bf Acknowledgments.}  This work was performed by by Lawrence Berkeley National Laboratory and was supported by the Director, Office of Energy Research, Office of High Energy and Nuclear Physics, Division of Nuclear Physics, of the U.S. Department of Energy under Contract DE-AC02-05CH11231.  This work also performed under the auspices of the U.S. Department of Energy by Lawrence Livermore National Laboratory under Contract DE-AC52-07NA27344.


\begin{thebibliography}{15}
\bibitem{bib1}  C. F. von Weizsacker, Z. Phys. A {\bf 96}, 431 (1935).
\bibitem{bib2}  V. M. Strutinsky, Nucl. Phys. {\bf A95}, 420 (1967). 
\bibitem{bib3}  W. D. Myers and W. J. Swiatecki, Nucl. Phys. {\bf 81}, 1 (1966).
\bibitem{bib15} K. Pomorski and J. Dudek, Phys. Rev. C {\bf 67}, 044316 (2003).
\bibitem{bib4}  G. Royer, Nucl. Phys. {\bf A807}, 105 (2008).
\bibitem{bib5}  M. W. Kirson, Nucl. Phys. {\bf A798}, 29 (2008).
\bibitem{bib11} S. Goriely, N. Chamel and J. M. Pearson, Phys. Rev. Lett. {\bf 102}, 152503 (2009).
\bibitem{bib12} P. Ring and P. Schuck, {\it The Nuclear Many-Body Problem} (Springer, New York, 2004).
\bibitem{bib9}  E. Wigner, Phys. Rev. {\bf 51}, 947 (1937).
\bibitem{bib14} W. D. Myers and W. J. Swiatecki, Ann. Phys.-New York {\bf 55}, 395 (1969).
\bibitem{bib10} R. C. Tolman, J. Chem. Phys. {\bf 17}, 333 (1949).
\bibitem{bib6}  R. Pohl, {\it et al.}, Nature {\bf 466}, 213 (2010).
\bibitem{bib7}  P. M\"oller, J. R. Nix, W. D. Myers and W. J. Swiatecki, At. Data Nucl. Data Tables {\bf 59}, 185 (1995).
\bibitem{bib8}  G. Audi, O. Bersillon, J. Blachot and A. H. Wapstra, Nucl. Phys. {\bf A729}, 3 (2003).
\bibitem{bib13} M. E. Fisher, Physics, {\bf 3} 255 (1967).

\end{thebibliography}
\end{document}